\documentclass[useAMS,usenatbib]{mn2e}
\usepackage{times}  
\usepackage{listings}
\usepackage{graphics}
\usepackage{subfigure}
\usepackage{graphicx}
\usepackage{graphicx,xspace}
\usepackage{amssymb}
\usepackage{amsmath}
\usepackage{aas_macros}
\usepackage{lscape}
\usepackage{rotating}
\usepackage{placeins}
\usepackage{natbib}
\usepackage{color}
     
\newcommand{\eagle}{{\sc{eagle}}\xspace}
\newcommand{\gadget}{{\sc{gadget-3}}\xspace}

\setcitestyle{notesep={ },round,aysep={},yysep={}}

\title[Scatter in MZR and gas flows]
{Explaining the scatter in the galaxy mass-metallicity relation with gas flows}

\author[M. L. van Loon et al.]{
\newauthor Maria L. van Loon\thanks{\rm E-mail: mvanloon98@gmail.com}$^{1}$,
Peter D. Mitchell$^{1}$
and Joop Schaye$^{1}$
\\
$^{1}$Leiden Observatory, Leiden University, P.O. Box 9513, 2300 RA Leiden, the Netherlands\\
}

\begin{document}
\date{\today}
\pagerange{\pageref{firstpage}--\pageref{lastpage}} \pubyear{2021}
\maketitle
\label{firstpage}


\begin{abstract}
The physical origin of the scatter in the relation between galaxy stellar mass and the metallicity of the interstellar medium, i.e. the Mass-Metallicity Relation (MZR), reflects the relative importance of key processes in galaxy evolution. The \eagle cosmological hydrodynamical simulation is used to investigate the correlations between the residuals of the MZR and the residuals of the relations between stellar mass and, respectively, specific inflow, outflow and star formation rate as well as the gas fraction for central galaxies. At low redshift, all these residuals are found to be anti-correlated  with the residuals of the MZR for $M_\star/\mathrm{M}_\odot \lesssim 10^{10}$.
The correlations between the residuals of the MZR and the residuals of the other relations with mass are interrelated, but we find that gas fraction, specific inflow rate and specific outflow rate all have at least some independent influence on the scatter of the MZR. 
We find that, while for $M_\star/\mathrm{M}_\odot > 10^{10.4}$ the specific mass of the nuclear black hole is most important, for $M_\star/\mathrm{M}_\odot \lesssim 10^{10.3}$ gas fraction and specific inflow rate are the variables that correlate most strongly with the MZR scatter.
The timescales involved in the residual correlations and the time that galaxies stay above the MZR are revealed to be a few Gyr. However, most galaxies that are below the MZR at $z=0$ have been below the MZR throughout their lifetimes.
\end{abstract}



\section{Introduction}
\label{Introduction}

The processes involved in galaxy formation and evolution are coupled and take place
over a range of length and time scales, which makes it challenging to develop
a full understanding. Hydrodynamical simulations are an established tool
  to assist in this, and modern simulations of cosmological
  volumes now reproduce many observational constraints well enough
  \cite[e.g.][]{Vogelsberger14,Schaye15,Dave16,2016Dobois}
  that they can be used with confidence to study aspects of galaxy evolution that
  are not readily observable, such as galactic gas accretion.

Observationally, such processes can be linked indirectly to quantities
  that are observable, such as the metallicity of the insterstellar medium (ISM).
  Gas-phase metallicity is influenced by processes that create or deplete heavy
  elements within galaxies, and by processes that increase or decrease the total
  gas content of galaxies \cite[see][for a recent review]{Maiolino_2019}.
The metallicity of the ISM is therefore influenced
by stellar evolution via the distribution of the metals created inside
  stars, by galactic outflows
  that carry metals and gas away from the ISM, and by gas accretion onto the
ISM, which dilutes its metal abundance assuming that the surrounding circum-galactic medium (CGM) is comparatively metal poor.

It is well established observationally that galaxy gas-phase metallicity
  correlates with galaxy stellar mass \cite[e.g.][]{Tremonti04,Zahid14,Curti20}, with
  a scaling relationship that is generally referred to as the Mass-Metallicity
  Relation (MZR). The MZR has a positive slope that is approximately constant for
$M_\star/\mathrm{M}_\odot \lesssim10^{10.5}$ and flattens for higher masses. It is
  usually accepted that the positive slope of the MZR for lower stellar mass galaxies is determined
by the higher efficiency of galactic outflows at lower masses,
due to their shallower gravitational potentials \cite[e.g.][]{Larson74}.

The scatter in the MZR is 0.1-0.2 dex, which is more than expected from measurement errors
\cite[][]{Maiolino_2019}, {and understanding this scatter offers the prospect
of advancing in turn our understanding of the underlying network of inflows,
outflows, and star formation within galaxies \cite[e.g.][]{finlatordave98, 2013dayal, Forbes14,  Guo_2016, LaraLopez19, Wang20, 2020delucia}.
Observationally, it has been reported that metallicity anti-correlates with star formation
rate or specific star formation rate at fixed stellar mass \cite[e.g.][]{Ellison08, Mannucci10, Curti20},
possibly forming a ``fundamental'' relation between the three quantities that is invariant
with redshift \cite[][]{LaraLopez10,Mannucci10}. Other observational studies do not find a correlation
between star formation rate and residuals in the MZR relation \cite[][]{Sanchez13, Sanchez19}, though
\cite{Cresci19} find that a significant correlation does exist for the same data if instead
residuals in both star formation rate and metallicity are compared at fixed stellar mass. 
Correlated residuals of the MZR with atomic gas mass or gas fraction \cite[][]{Bothwell13} and for molecular hydrogen mass \cite[][]{Bothwell16} have also been reported.
Generally speaking, theoretical models and cosmological simulations predict qualitatively similar
anti-correlations of metallicity with star formation rate or gas mass at fixed stellar mass
\cite[e.g.][]{Yates12,Lilly13,Lagos16,De_Rossi_2017,Torrey19,deLucia20}.

Theoretical work has also been used to study the connection betwen the MZR and variables
that are not readily observable. For example, using simple idealised galaxy evolution models,
\cite{Forbes14} and \cite{Wang20} study how fluctuations in gas inflows rates can drive
the MZR scatter, emphasing the importance of the relationship between the fluctuation
timescale for inflows and the gas consumption timescale in the ISM.
\cite{deLucia20} use a more complex semi-analytic model of galaxy formation to
study the physical drivers of the MZR relation, and arrive at the conclusion that gas
accretion drives the MZR scatter, with galaxies below the MZR relation being driven
by sustained periods of gaseous inflow. Using hydrodynamical simulations of a cosmological
volume, \cite{Torrey19} arrive at similar conclusions by studying how galaxies evolve
in metallicity, gas mass, and star formation rate over finite time intervals.

Here, we extend this body of work by using cosmological hydrodynamical simulations of
a representative volume to 
directly study the connection between the MZR and inflow rates, outflow rates, and
star formation rates. Specifically, we use the \eagle simulations,
building on previous analyses of the MZR in these simulations by \cite{Schaye15}, \cite{crain15}, \cite{Lagos16}, \cite{De_Rossi_2017}, \cite{Almeida_2018}, \cite{trayford19}
and \cite{Zenocratti20}.
Our work complements that of more idealised modelling that
generally infers the properties of inflows and outflows based on observables, and
also complements semi-analytic modelling by self-consistently allowing for potential coupling between
outflows and inflows \cite[i.e. ``preventative feedback'', e.g.][]{2002mo_mao, 2011voort, Dave12}.

This study is organised as follows. Section \ref{Methods} briefly describes the \eagle
simulations and how the various variables used in this study have been measured and defined.
Section \ref{Results} presents the results of the study, including some individual
examples, and the correlations between the residuals of the MZR and residuals of the relations between 
stellar mass and the other variables (specific inflow, outflow and star formation rates and the gas fraction). We explore the stellar
mass dependence (section \ref{stellar_mass_dependence}), inter-variable coupling (section \ref{Coupling}),
the contribution to the MZR scatter (section \ref{Explained_Scatter}), and the relevant timescales
(section \ref{Timescales}). Section \ref{Summary} provides
a summary of the main results. 

\section{Methods}
\label{Methods}
We use data from the \eagle simulation project, which consists of a suite of cosmological hydrodynamical simulations \cite[][]{Schaye15, crain15}, which have been made public \cite[][]{2016McAlpine}. \eagle simulates representative cosmological volumes, including gas, stars, supermassive black holes and dark matter, using smoothed particle hydrodynamics (SPH) to solve the equations of hydrodynamics with a modified version of the \gadget code \citetext{last described in \citealp{2005Springel}; see \citealp{2015schaller} for the employed formulation of SPH.}
The simulation assumes a $\Lambda$CDM cosmological model and uses 'subgrid' models for relevant unresolved physics. The parameters for the subgrid models governing stellar and AGN feedback were calibrated to fit the observed galaxy stellar mass function at $z\approx0$, as well as the local disk galaxy size - stellar mass relation and the local black hole mass - stellar mass relation.

The \eagle simulation suite consists of a set of “Reference” simulations with a common set of model parameters, as well as a number of variations. All \eagle measurements in this article are taken from the largest Reference model simulation, Ref-L100N1504, which simulates a $(100 \,\mathrm{Mpc})^3$ box volume with $1504^3$ dark matter and $1504^3$ baryonic particles. The (initial) particle masses in this simulation are $1.8\times10^6 \mathrm{M}_\odot$ for gas and stars, and $9.7\times10^6 \mathrm{M}_\odot$ for dark matter.
Note that we opt to use this simulation over the smaller volume "Recal" simulation since our analysis involves multi-dimensional binning of galaxies across several variables,
which benefits greatly from the factor $8^3$ larger volume of the main reference simulation.
The Recal simulation uses eight times better mass resolution than the main reference simulation, uses re-calibrated model parameters, and better reproduces 
the shape of the observationally inferred MZR at low stellar mass \cite[][]{Schaye15,De_Rossi_2017}.
We compare our main results between a standard \eagle resolution simulation with fiducial parameters against Recal in Appendix~\ref{ap_recal}, and
find that our results are qualitatively the same, with minor quantitative differences.

Since \eagle employs SPH, it is possible to track simulation particles across time. We use this information to measure galactic inflow and outflow rates by determining how much gas enters or leaves the ISM of a given galaxy between two simulation snapshots. The precise method used to measure inflow and outflow rates is described in detail in \cite{mitchell2020galactic, Mitchell_2020}. We define the ISM as the combination of star-forming gas \cite[meaning its density exceeds the metallicity-dependent threshold introduced in][]{Schaye_2004}, and also non-star forming gas with density $n_{\mathrm{H}}>0.01 \, \mathrm{cm}^{-3}$ that is located within 20\% of the halo virial radius, approximately matching a selection of neutral atomic hydrogen \cite[e.g.][]{2013Rahmati}.

To account for particles fluctuating across the ISM boundary, outflowing gas particles are selected by their radial velocity, ensuring that they have moved a significant distance. The instantaneous radial velocity and the difference in radius between two subsequent snapshots, divided by the time elapsed between them, must both exceed 0.25 times the maximum circular velocity of the halo. Particles that leave the ISM without meeting these criteria are tracked through subsequent snapshots until they: \textit{a)} meet the outflow criteria, or \textit{b)} have re-entered the ISM (at which point they are not counted towards the inflow rate), or \textit{c)} until three halo dynamical times have passed.

Inflowing gas particles can be distinguished into three different types, as presented by \cite{mitchell2020galactic}. 
First, particles can be accreted onto a galaxy for the first time. 
Second, particles can be recycled from a progenitor of the current galaxy. 
Last, particles can be transferred from the ISM of another galaxy.
These categories are added together to create the total inflow onto a galaxy, which is the variable used in this study. 
Note that a distinction is made between inflows and gas accretion due to galaxy mergers; we choose not to include the latter in our analysis (defining mergers as satellites with a maximum past halo mass greater than $9.7 \times 10^8 \, \mathrm{M_\odot}$, corresponding to 100 dark matter particles at fiducial \eagle resolution).

Inflow, outflow, and star formation rates are measured using adjacent snapshots in a 200 snapshot grid. 
Inflowing (outflowing) gas mass, or formed stellar mass between snapshots is divided by the time elapsed between snapshots, 
which is $\approx 120$ Myr at low redshift and becomes smaller at higher redshifts \cite[for the exact temporal spacing, see figure A1 in][]{Mitchell_2020}. 
In parts of our analysis, the temporal spacing is adjusted by tracking galaxies across snapshots and integrating their inflow, outflow or star formation 
rates over longer timescales, which we fix to be a constant fraction of the halo dynamical time.


This study focusses only on central galaxies, thereby simplifying the interpretation of the results by excluding environmental effects. Furthermore, galaxies with zero ISM gas particles are excluded due to their undefined metallicity. Similarly, for figures showing data concerning inflows, galaxies with zero inflow rate have been excluded (the same holds for outflows and star formation rate (SFR)). Apart from this selection, we select galaxies only by stellar mass and redshift. However, most figures show (with transparent and dashed lines) where the interpretation becomes uncertain due to a majority of galaxies ($>50\%$) containing $<$ 20 inflowing, outflowing, newly formed star particles, or $<$ 200 star forming gas particles. Furthermore, we discard data points including fewer than 10 galaxies, or where $>50\%$ of galaxies have $<$  10 inflowing, outflowing, newly formed star particles, or $<$ 100 star forming gas particles, to reduce the noise in the plots. None of these choices strongly influence the results.

\section{Results}
\label{Results}

\begin{figure}
\includegraphics[width=20pc]{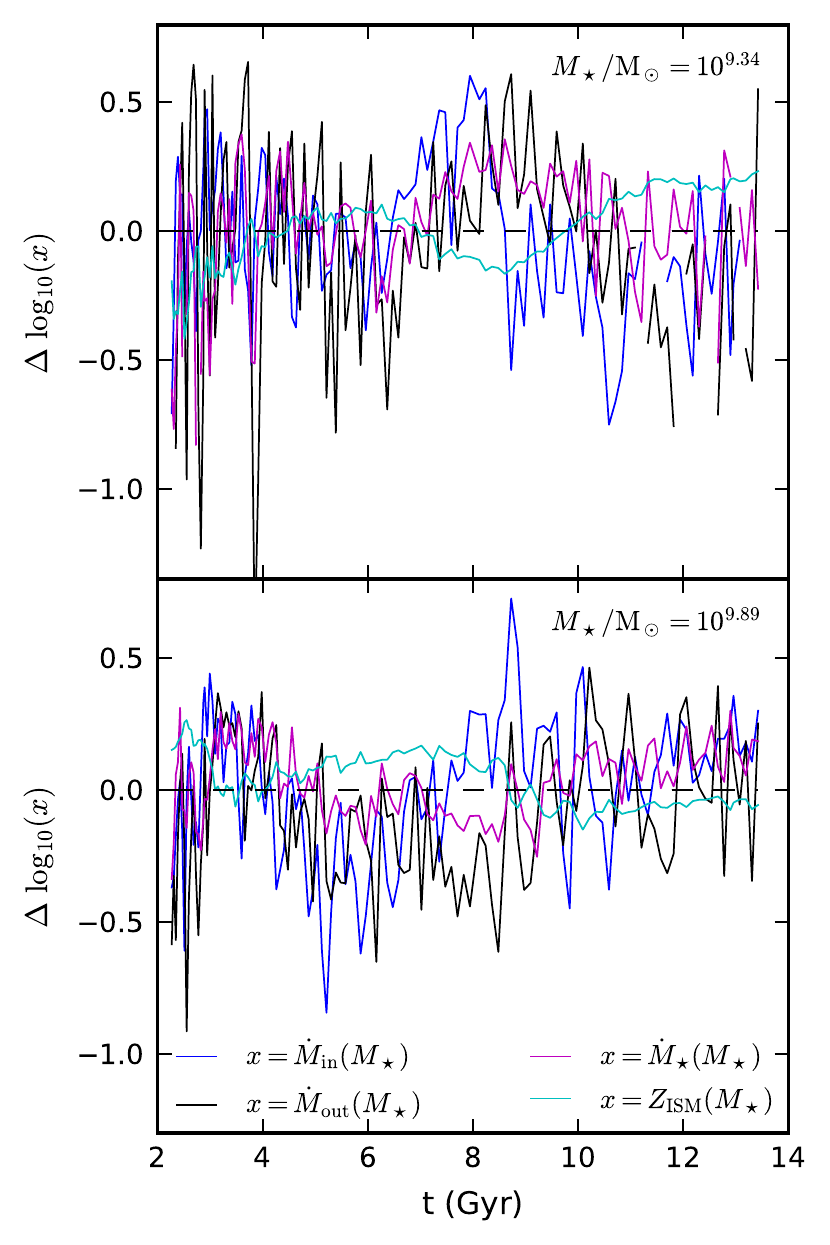}
\caption{Examples of the residual inflow rate ($\dot{M}_{\mathrm{in}}$), outflow rate ($\dot{M}_{\mathrm{out}}$), star formation rate ($\dot{M}_\star$)
and metallicity of the interstellar medium ($Z_{\mathrm{ISM}}$) of two individual galaxies tracked through time, with present-day masses $M_\star/\mathrm{M_\odot} = 10^{9.34}$ (top) and $M_\star/\mathrm{M_\odot} = 10^{9.89}$ (bottom). 
At $z=0$, samples of galaxies are chosen based on stellar mass (Top: $M_\star/\mathrm{M_\odot} = 10^{9.4\pm0.3}$; Bottom: $M_\star/\mathrm{M}_\odot = 10^{10.0\pm0.3}$). 
Residuals are taken with respect to the medians for these same samples at each snapshot.
The inflow, outflow and star formation rate show fluctuations across a wide range of timescales. Metallicity anti-correlates with these variations but it fluctuates less strongly.
}
\label{fig_ind}
\end{figure}

Fig.~\ref{fig_ind} shows two examples of typical star forming galaxies, whose inflow rate, outflow rate, star formation rate and metallicity have been tracked though time. 
At $z=0$ samples of the galaxies are created based on their stellar mass (bin width = 0.3 dex). 
These samples are then used to determine the median values in each snapshot, with respect to which the residuals of the variables are calculated. 
Fig.~\ref{fig_ind} plots the evolution of there residuals for two individual galaxies. It illustrates clearly that when the inflow, outflow and star formation rates are higher  than the median of the sample (i.e. the residual is positive), the metallicity tends to be lower (i.e. the residual is negative) and vice versa. 
The variables fluctuate over a range of timescales spanning $\approx 0.25$ Gyr to $\approx 5$ Gyr, where the longer timescale fluctuations 
show a clear anti-correlation between metallicity and the other variables. 
Inflow rates (blue line) generally seem to be the first to rise or fall, after which other variables react, if these (anti-) correlations are interpreted causally. 
The short timescale fluctuations of inflow, outflow and star formation rate are large compared to those in the metallicity. 

Fig.~\ref{fig_mb} shows the median residuals of the relations between stellar mass and, respectively, the specific inflow rate, specific outflow rate, 
gas fraction and specific star formation rate plotted against the residuals of the mass-metallicity relation for different stellar 
mass bins ($M_\star/\mathrm{M_\odot} = 10^{9.0 \pm 0.3}$ (blue), $M_\star/\mathrm{M_\odot} = 10^{10.0 \pm 0.3}$ (green), $M_\star/\mathrm{M_\odot} = 10^{11.0 \pm 0.3}$ (magenta)), at $z<0.1$. 
The 'specific' in specific inflow/outflow/star formation rate means the variables are divided by the stellar mass of the galaxy, thus 
factoring out any residual stellar mass dependence remaining within the finite stellar mass bins.

Focusing first on the stellar mass bin $M_\star/\mathrm{M_\odot} = 10^{10.0 \pm 0.3}$, the residual specific inflow rate, specific outflow rate, gas fraction and specific star formation rate all anti-correlate with the residual metallicity (see Fig. \ref{fig_mb}, green lines in the top left, top right, bottom left, bottom right panels, respectively). 
When interpreted causally, this suggests that the processes responsible for increasing these variables generally lower the gas metallicity in a galaxy, with a possible lower limit to the effect (the correlations flatten at extremely low values). 
Due to the coupling of the variables however (see sections \ref{Introduction} and \ref{Coupling}), 
the correlations need to be investigated further before understanding which physical processes cause them. 

\begin{figure*}
  \begin{center}
\includegraphics[width=40pc]{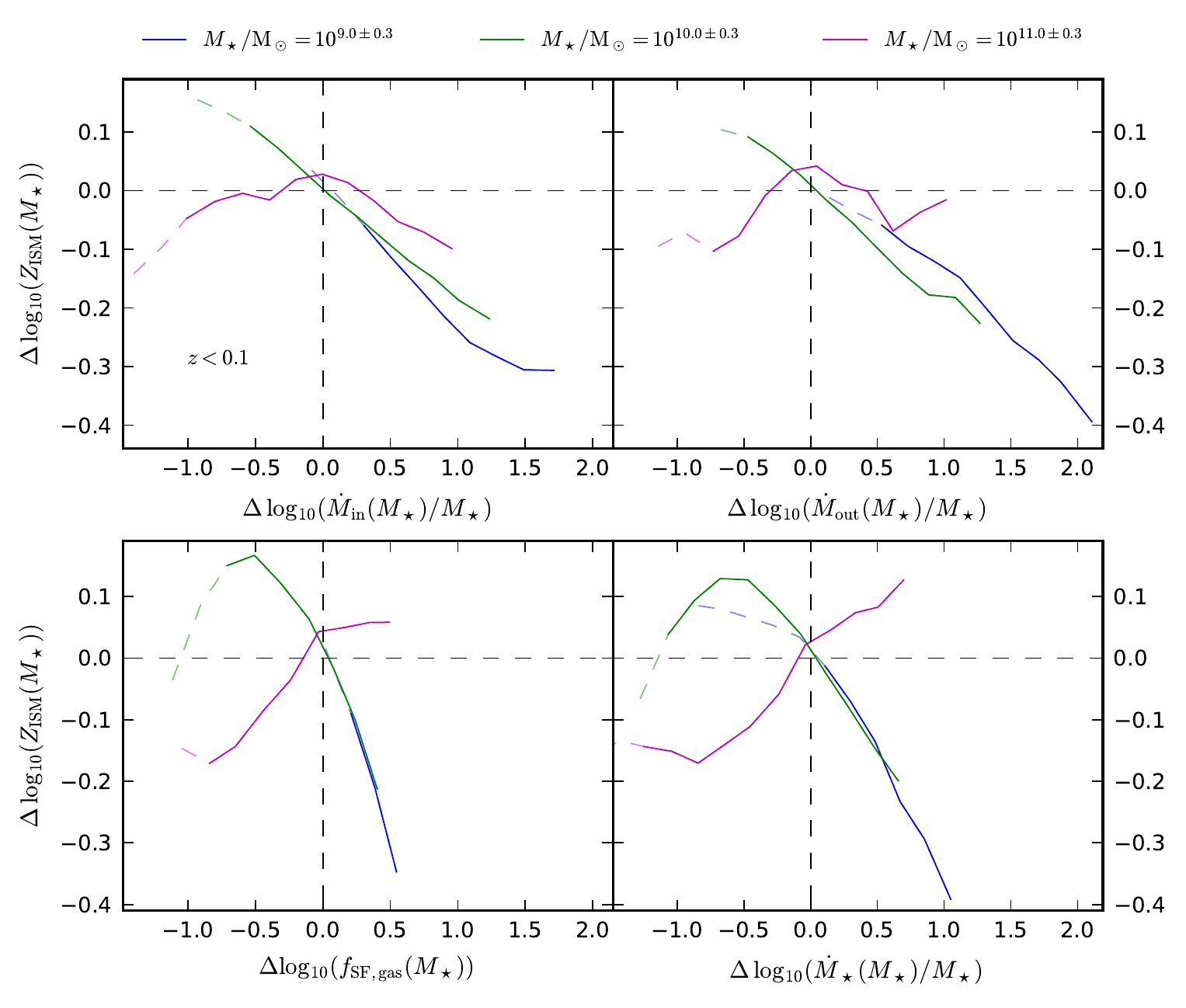}
\caption{The relations between the median residual metallicity of the interstellar medium and  residual specific inflow rate (top left), residual specific outflow rate (top right), residual gas fraction (bottom left) and residual sSFR (bottom right) for central galaxies with $M_\star/\mathrm{M_\odot} = 10^{9.0 \pm 0.3}$ (blue lines), $M_\star/\mathrm{M_\odot} = 10^{10.0 \pm 0.3}$ (green lines), $M_\star/\mathrm{M_\odot} = 10^{11.0 \pm 0.3}$ (magenta lines), at $z<0.1$.
The residuals are taken w.r.t. the medians as a function of stellar mass.
A dashed and transparent line style indicates regions where more than 50\% of galaxies in the bin contain $<$ 20 inflowing (top left panel), outflowing (top right panel), newly formed star particles (bottom right panel) or $<$ 200 star forming gas particles (bottom left panel).
The lines are cut off when more than 50\% of galaxies in the bin contain $<$ 10 inflowing (top left panel), outflowing (top right panel), newly formed star particles (bottom right panel), or $<$ 100 star forming gas particles (bottom left panel). or when bins contain fewer than 10 galaxies.
For stellar masses $<10^{10.3} \mathrm{M_\odot}$, the shapes and slopes are comparable across the residual correlations. For stellar masses $>10^{10.7} \mathrm{M_\odot}$, they deviate, forming two groups: specific inflow and outflow rate (top), and gas fraction and sSFR (bottom).
}
\label{fig_mb}
  \end{center}
\end{figure*}

A starting point for uncovering the physical origin of the MZR scatter can be a comparison between the correlations shown in Fig. 2. 
A comparison of slopes reveals that the correlations of the residual metallicity with the residual specific inflow and outflow rate match closely in steepness and shape, 
whereas the residual correlations with the gas fraction and sSFR residual correlations are much steeper, the gas fraction residual correlation being the steepest.
This supports the idea that the correlation with the gas fraction may be caused by a combination of the influence of multiple other variables.
Note that unlike the other variables considered, gas fraction is dimensionless and is not defined in terms of a timescale, and any comparison of
slopes should be interpreted with this in mind.

\subsection{Stellar mass dependence}
\label{stellar_mass_dependence}

Fig.~\ref{fig_mb} depicts the residual correlations for three stellar mass bins, 
of $M_\star/\mathrm{M_\odot} = 10^{9.0 \pm 0.3}$, $10^{10.0 \pm 0.3}$ and $10^{11.0 \pm 0.3}$, at $z<0.1$. 
The redshift evolution of the residual correlations has also been investigated and is found to be small, as can be viewed in Appendix \ref{Redshift_evolution}. 
Considering the stellar mass dependence of the residual correlations, Fig.~\ref{fig_mb} shows that the lower 
($M_\star/\mathrm{M_\odot} = 10^{9.0 \pm 0.3}$, blue lines in Fig.~\ref{fig_mb}) and intermediate 
($M_\star/\mathrm{M_\odot} = 10^{10.0 \pm 0.3}$, green lines in Fig.~\ref{fig_mb}) stellar mass bins 
of all four residual correlations show a comparable progression, varying only slightly in shape and slope. 

The higher stellar mass bin ($M_\star/\mathrm{M_\odot} = 10^{11.0 \pm 0.3}$) deviates in shape from the lower two,
especially for the residual trends with gas fraction and star formation rate, which change in sign to
clear positive correlations, consistent with the previous analysis of \eagle by \cite{De_Rossi_2017} \cite[see also][for a similar result from a semi-analytic model]{Yates12}.
Observationally, \cite{Yates12} find the same trend in Sloan Digital Sky Survey (SDSS) data, with star formation rates
inverting from a negative to a positive correlation with metallicity at fixed stellar mass above the characteristic
stellar mass $\sim 10^{10.2} \, \mathrm{M_\odot}$. \cite{Mannucci10} find that
metallicity is only very weakly dependent on star formation rate at fixed stellar mass in
high-mass galaxies. \cite{Curti20} have since demonstrated that these results depend sensitively
on whether SDSS fiber aperture corrections are applied to star formation rates, and on which
metallicity calibrator is used, concluding that correlations with star formation rate are inconclusive
at high stellar mass for the SDSS data (and that such a correlation inversion does not
seem to appear in other datasets). Given the theoretical prediction that there is a significant inversion
in the correlation between star formation rate and metallicity at high
stellar mass, this is clearly an interesting avenue to continue exploring in future observational work.
In contrast to the residual correlations with star formation rate and gas fraction, the residual correlations with inflow and 
outflow rate deviate in a more complex manner at high stellar mass, appearing to show
a positive correlation for negative x-axis residuals, but retaining an overall anti-correlation for positive
x-axis residuals.

By comparing \eagle simulations with and without AGN feedback, \cite{De_Rossi_2017} find that the inversion
of the dependence on the gas fraction and star formation rate residuals at high stellar mass can be attributed to the AGN feedback. We explore this aspect further in Fig.~\ref{fig_mbh}, which shows the correlation 
between residual ISM metallicity and residual specific black hole mass. Fig.~\ref{fig_mbh} supports the idea that for $M_\star/\mathrm{M}_\odot>10^{10.7}$ the shapes 
of the residual correlations (Fig.~\ref{fig_mb}) are likely caused by 
the influence of AGN feedback on larger mass galaxies. For the highest mass bin, the correlation 
between residual metallicity and residual specific black hole mass is steeper and negative, suggesting that high black hole 
masses correlate strongly with a low metallicity. For the lower stellar mass bins,
the residual correlations are almost flat until extreme positive residuals in black hole mass, indicating
a possible causal connection whereby black holes above a threshold mass are able to influence the metallicity
of the galaxy.

\begin{figure}
\includegraphics[width=20pc]{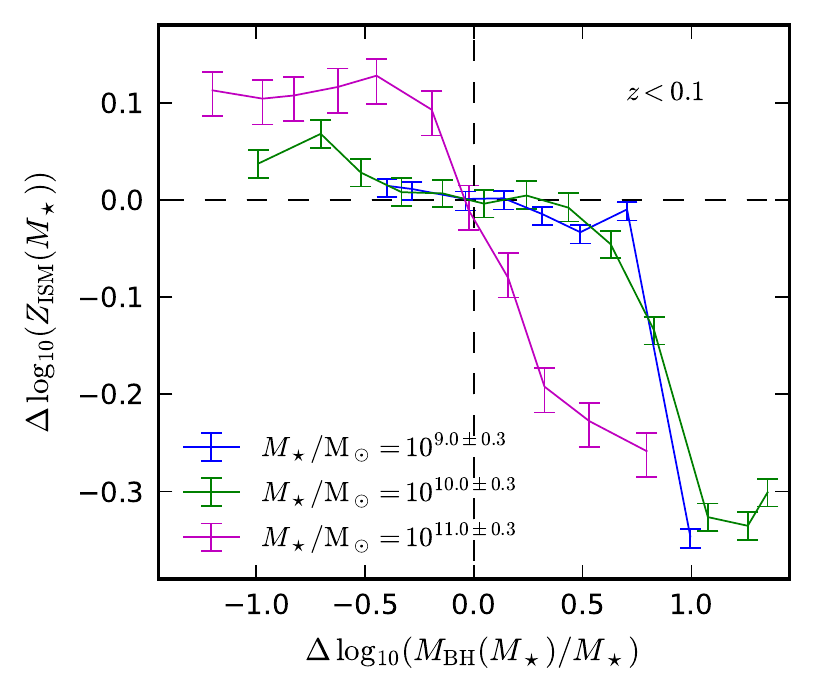}
\caption{The median residual of the relation between ISM metallicity and stellar mass as a function of the residual between specific black hole mass and stellar mass for central galaxies with $M_\star/\mathrm{M}_\odot = 10^{9.0 \pm 0.3}$ (blue line), $M_\star/\mathrm{M}_\odot = 10^{10.0 \pm 0.3}$ (green line), $M_\star/\mathrm{M}_\odot = 10^{11.0 \pm 0.3}$ (magenta line), at $z<0.1$.
Only bins containing more than 10 galaxies are shown.
Error bars depict the $16^{th}$ and $84^{th}$ percentiles of the distribution of medians obtained by bootstrapping our data,
implying in particular that the downturn in residual metallicity at high residual black hole mass is statistically significant.
Residual metallicity and residual specific black hole mass are anti-correlated in the highest mass bin for black hole masses, 
and are anti-correlated in the lower-mass bins for galaxies with unusually large black hole masses ($\Delta \log_{10}(M_{\mathrm{BH}}(M_\star)/M_\star)>0.75$ dex).
}
\label{fig_mbh}
\end{figure}

\subsection{Coupling between different variables}
\label{Coupling}

While Fig.~\ref{fig_mb} shows clear correlations between metallicity and inflow, outflow, and star formation rates, and
also gas fraction, the latter four variables can of course also be correlated with each other, rather than each directly
driving a causal connection.
One way to probe the coupling of variables, is by correcting the residual correlations for others, thus making it possible to judge the influence of each variable on the other residual correlations. By comparing all possibilities, insight is gained into which variable may be most fundamental. 
In Fig.~\ref{fig_coup}, each panel shows the residual correlation of one of the four variables (black lines, $M_\star/\mathrm{M}_\odot = 10^{10.0 \pm 0.3}, z<0.1$; there curves are identical to the green lines in Fig. \ref{fig_mb}), 
and the residual correlation after correcting for one of the other variables 
(blue: specific inflow rate, green: specific outflow rate, magenta: specific star formation rate, cyan: gas fraction).

The correction is calculated by taking the residuals of the residual correlations. First, the median residuals of the relations between stellar mass and, respectively, metallicity and (e.g.) specific outflow rate are calculated (respectively $\Delta \, \log_{10}( Z_{\mathrm{ISM}} (M_\star) )$ and $\Delta\, \log_{10} (\dot{M}_{\mathrm{out}} (M_\star) / M_\star )$, which are plotted in Fig. 2). Then, the residuals of the relation between these two variables are calculated, creating a new variable:
\begin{equation*}
    \Delta \, \left( \Delta\,\log_{10}( Z_{\mathrm{ISM}} (M_\star) ) \left[ \Delta\, \log_{10} (\dot{M}_{\mathrm{out}} (M_\star) / M_\star ) \right] \right)
\end{equation*}
which is then plotted against the residuals of the relation between stellar mass and another variable (e.g. specific inflow rate) to create Fig. 4.

\begin{figure*}
  \begin{center}
\includegraphics[width=40pc]{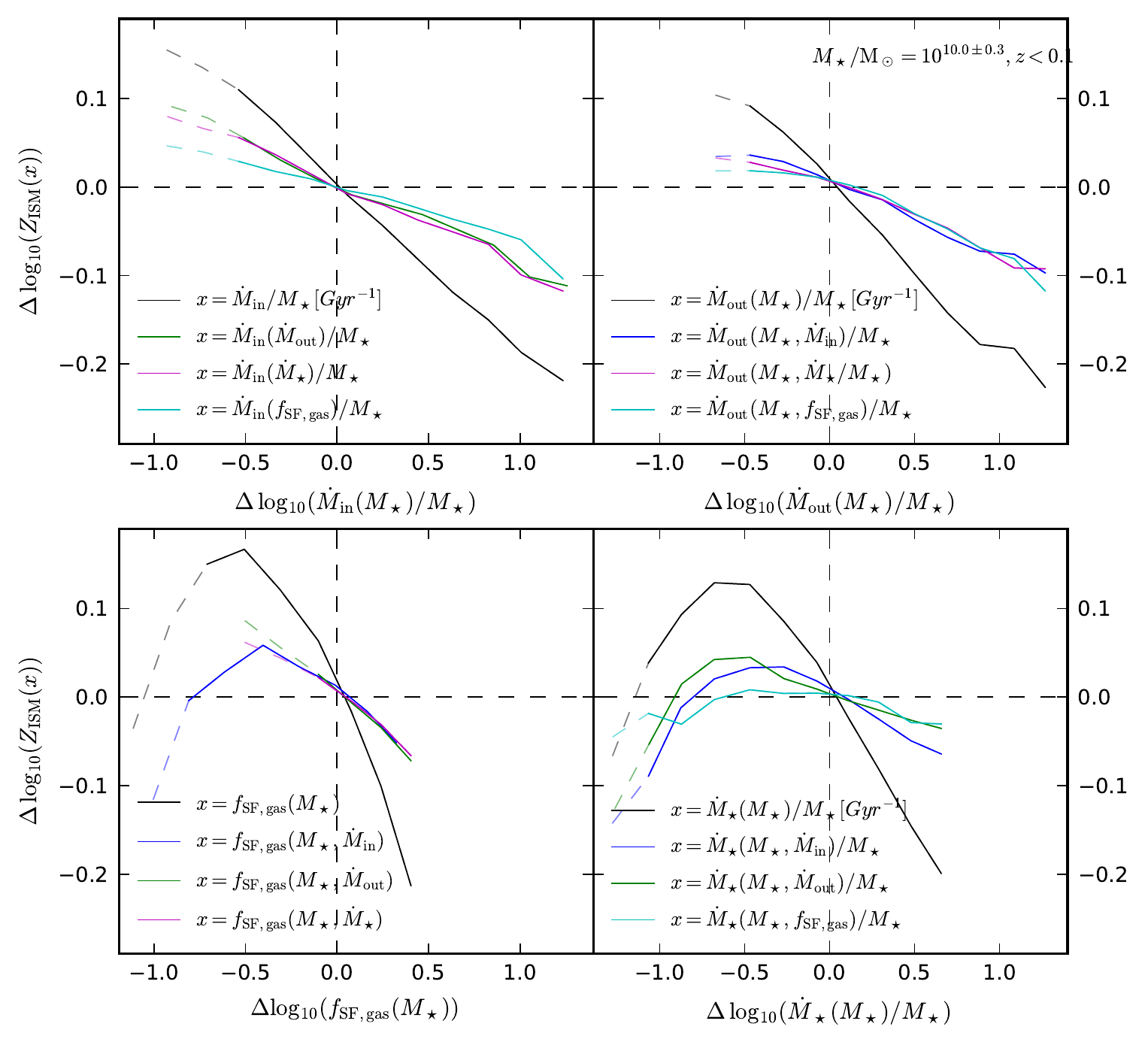}
\caption{The median residual of the gas mass-metallicity relation as a function of the residual specific inflow rate (top left), specific outflow rate (top right), gas fraction (top left), and sSFR (bottom right) for central galaxies with $M_\star/\mathrm{M_\odot} = 10^{10.0 \pm 0.3}$, at $z<0.1$.
\textit{Black:} The residuals of the relations with stellar mass.
\textit{Blue:} The residual correlation corrected for the relation between residual metallicity and residual specific inflow rate.
\textit{Green:} The residual correlation corrected for the relation between residual metallicity and residual specific outflow rate.
\textit{Magenta:} The residual correlation corrected for the relation between residual metallicity and residual specific star formation rate.
\textit{Cyan:} The residual correlation corrected for the relation between residual metallicity and residual the gas fraction.
Line styles are as in Fig.2.
Each residual correlation flattens significantly when corrected for the dependence of metallicity on one of the other variables implying that all variables are coupled. 
}
\label{fig_coup}
  \end{center}
\end{figure*}

Fig.~\ref{fig_coup} shows that each residual relation flattens considerably after correcting for another residual correlation (compare the coloured lines to the black lines), 
indicating that none of the variables are fully independent.
When studying the bottom right panel in Fig.~\ref{fig_coup}, it becomes clear that the effect of the sSFR on the residual metallicity is completely determined by the 
gas fraction (cyan line) and to a lesser extent by the specific inflow and outflow rates (blue and green lines, respectively), 
with the exception of extreme values ($\Delta \, \log_{10}(\mathrm{sSFR}) <-0.7, \,\,\Delta \, \log_{10}(\mathrm{sSFR}) >0.3$). 
The relations between residual metallicity and, respectively, residual specific inflow rate, residual specific outflow rate and residual gas fraction flatten considerably when corrected for the other variables. However, there is no significant difference between them.


Combining all the information in Fig.~\ref{fig_coup}, we can conclude that the gas fraction, specific inflow, outflow and star formation rate all play an independent role in determining the metallicity of a galaxy, but that they are also coupled. The independent influence of star formation rate is least important.
The gas fraction can be seen physically as a time integrated aggregate of the three other processes, and will therefore naturally correlate strongly with the metallicity and affect other correlations. 

\subsection{The scatter in the MZR}
\label{Explained_Scatter}

We now consider directly how much of the MZR scatter can be explained by each of the variables considered, using a similar
method as used by \cite{Matthee_2019} to explore the origin of the scatter in star formation rates at fixed stellar mass in the \eagle simulation.
As a potentially important caveat, the \eagle simulation has $\approx0.13$ dex scatter in the MZR, which is slightly larger than the $\approx 0.1$ dex
scatter in the MZR reported using SDSS observations by \cite{Tremonti04}. The metallicities presented here are measured differently from those in observational work, which
are generally limited to tracing gas in HII regions. These differences may explain the difference in scatter, but it should nonetheless be kept in mind that the 
MZR scatter in \eagle may indeed be overestimated.

Fig.~\ref{fig_expl_sc} shows how much of the original MZR scatter (black line) remains after correcting for the residual correlation with another variable, using the procedure described in section \ref{Coupling}. 
Fig.~\ref{fig_expl_sc} then depicts the remaining (1$\sigma$) scatter. The residual correlations were recalculated for different stellar mass bins (width = 0.15 dex) to account for the stellar mass dependencies of the residual correlations and include only central galaxies at $z<0.1$.

\begin{figure}
\includegraphics[width=20pc]{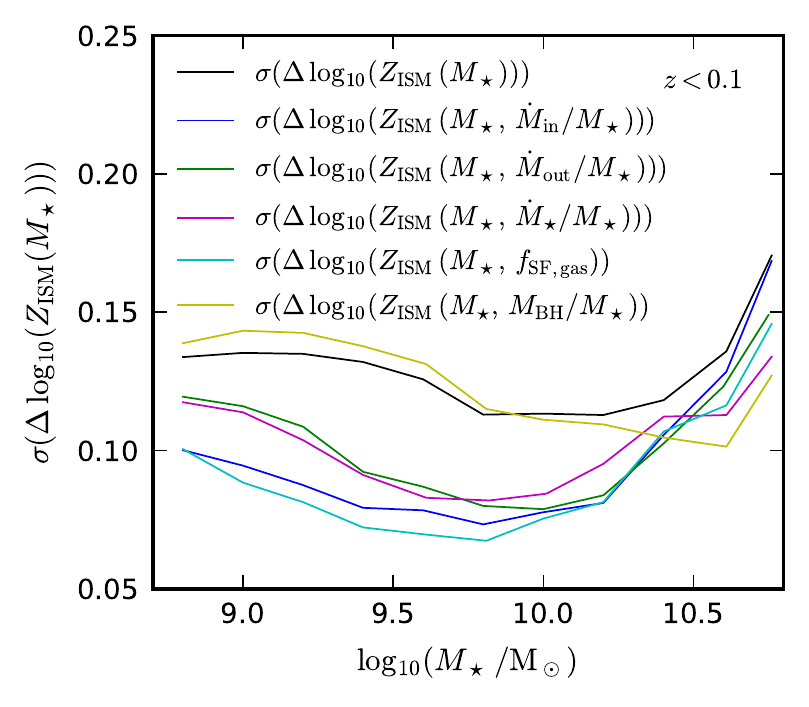}
\caption{The mass dependence of the (1$\sigma$) scatter in the median MZR (black), when corrected for the correlations between residual metallicity and residual specific inflow rate (blue), residual specific outflow rate (green), residual sSFR (magenta), residual gas fraction (cyan) and residual specific black hole mass (yellow) for central galaxies, at $z<0.1$ (where the residuals are taken w.r.t. the median relations with stellar mass).
At low stellar masses ($M_\star/\mathrm{M_\odot} < 10^{10.4}$), the gas fraction and specific inflow rate reduce the scatter the most. At high stellar masses ($M_\star/\mathrm{M_\odot} > 10^{10.4}$), the specific black hole mass reduces the scatter the most.
}
\label{fig_expl_sc}
\end{figure}

Fig.~\ref{fig_expl_sc} shows that up to $\approx0.06$ dex of the $\approx0.13$ dex scatter in the MZR can be explained by correcting for one of the variables (e.g. the gas fraction, cyan line). At low stellar masses ($M_\star/\mathrm{M}_\odot < 10^{10.4}$) correcting for the gas fraction (cyan line) or the specific inflow rate (blue line) reduces the scatter the most, marking these two variables as most influential. When correcting for the sSFR (magenta line) or specific outflow rate (green line) residual correlation, only $\approx0.025$ dex in scatter is explained at $M_\star/\mathrm{M}_\odot<10^{9.3}$, compared to the $\approx0.04$ dex at $M_\star/\mathrm{M}_\odot \approx10^{9.6}$. At $M_\star/\mathrm{M}_\odot > 10^{10}$ the amount of explained scatter diminishes for all variables, with the exception of specific black hole mass (yellow line), which only accounts for scatter at higher stellar masses and is the most important variable for $M_\star/\mathrm{M}_\odot > 10^{10.4}$. This can be explained by the effects of AGN feedback on the metallicity of high-mass galaxies.

Fig.~\ref{fig_expl_sc} provides strong evidence supporting the idea of inflows of gas having a larger influence on the scatter in the MZR than outflows or star formation for $M_\star/\mathrm{M_\odot} \lesssim 10^{10}$. The gas fraction reduces the scatter in the MZR the most, which supports the idea that the physical origin of the scatter lies in more than one variable, under the assumption that the gas fraction reflects the influence of multiple variables (inflow, outflow and star formation rates). Note that this does not exclude the possibility of a direct influence of the gas fraction on the metallicity (higher gas surface densities could change the effectiveness of feedback, for example). Finally, we note that after correcting for the secondary dependence on star formation rate, the
remaining MZR scatter in \eagle ($\approx 0.1$ dex) is larger than the $0.05 \, \mathrm{dex}$ scatter reported by, e.g., \cite{Mannucci10,Curti20} after star formation rates are
similarly corrected for in observational work, which (as with the total scatter at fixed stellar mass) may indicate that the scatter is too large in \eagle.

\subsection{Timescales}
\label{Timescales}

Fig.~\ref{fig_expl_sc} shows that not all of the scatter in the MZR can be explained by just one of the individual variables considered. 
We remind the reader however that until now we have presented inflow and outflow rates that are measured over a $\approx 120 \, \mathrm{Myr}$ time interval.
Recalling from Fig.~\ref{fig_ind} that the metallicity shows long timescale ($\gg1$ Gyr) fluctuations, we now 
investigate the timescales involved in the residual correlations by averaging the variables over longer time intervals.

\begin{figure*}
  \begin{center}
\includegraphics[width=40pc]{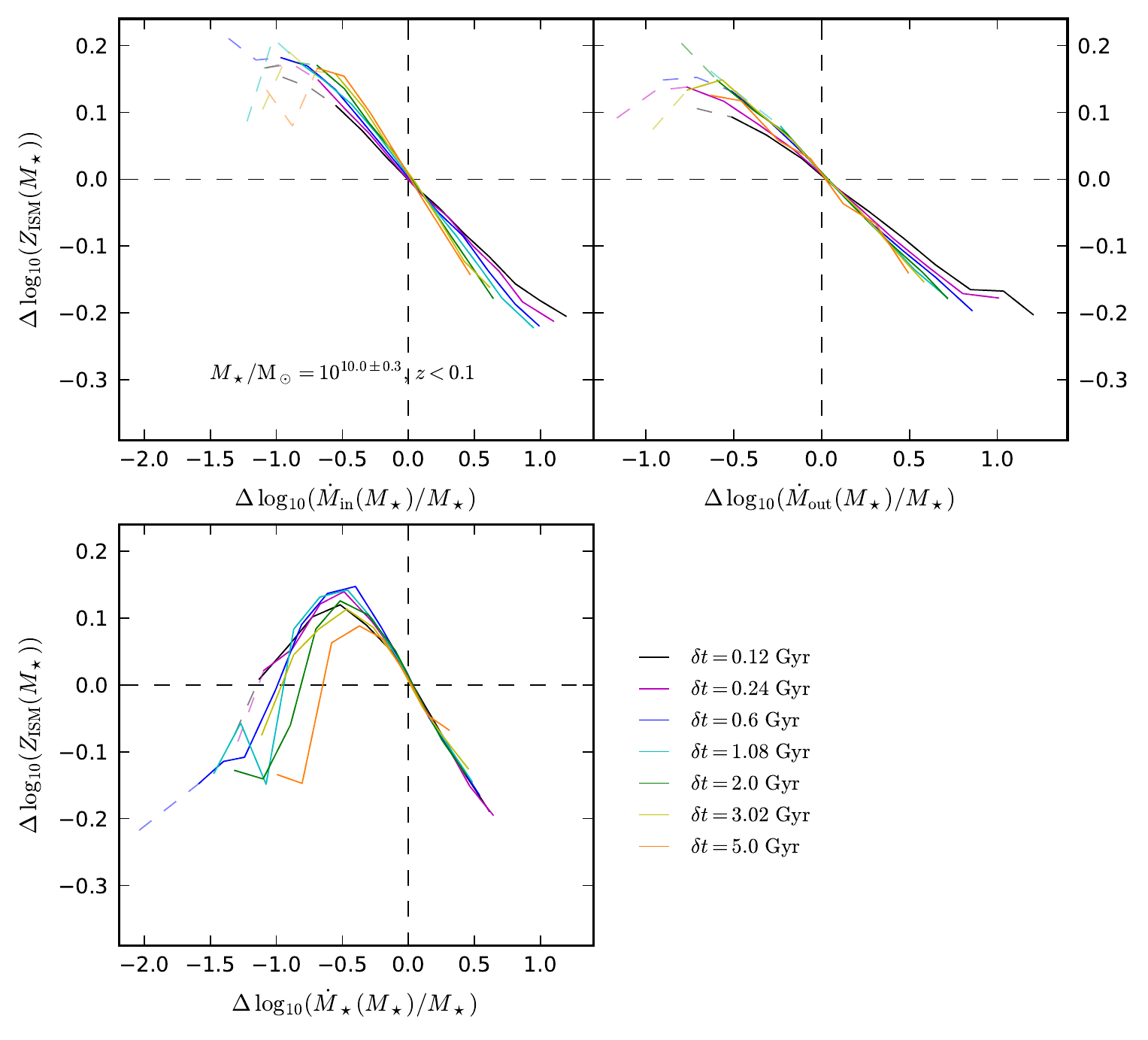}
\caption{The relations between the median residual of the median MZR and the residuals of the relations with stellar mass of, respectively, specific inflow rate (top left), specific outflow rate (top right) and sSFR (bottom left) for central galaxies with $M_\star/\mathrm{M}_\odot = 10^{10.0 \pm 0.3}$, at $z<0.1$. 
Different colours correspond to variables averaged over different time intervals.
Line styles are as in Fig.2.
The specific inflow and outflow rate residual relations steepen when the time interval increases, indicating the importance of longer timescales.
}
\label{fig_tint}
  \end{center}
\end{figure*}

Fig.~\ref{fig_tint} illustrates that the correlations with the residual specific inflow, outflow and star formation rate are likely dominated by fluctuations on long timescales. 
The figure shows the residual correlations for measurements of the variables over different time intervals, which smooth the fluctuations that occur on shorter timescales. 
Galaxies selected at $z<0.1$ are tracked back in time, across the desired time interval, and their inflows/outflows/formed stars are added together.

When the time interval increases, smoothing the shorter timescale fluctuations, the relations with the residual specific inflow and outflow rates steepen somewhat (top panels). This slight increase of steepness is due to the decrease of 'noise' (shorter timescale fluctuations in the variables), indicating that fluctuations on longer timescales are likely the main cause of the correlations. The residual correlation for specific outflow rate saturates at a time interval of $>1$ Gyr, suggesting that fluctuations on this timescale can no longer be  considered to be noise.
The inflow and outflow panels indicate that short timescales are not dominating the residual correlations, as the correlations would have flattened when the fluctuations over these timescales are smoothed.

The relation with the residual specific star formation rate does not change when measured over different time intervals, 
indicating that for this correlation, short timescales are not the cause. Why the residual correlation does not become steeper like the others is unclear.

\begin{figure}
\includegraphics[width=20pc]{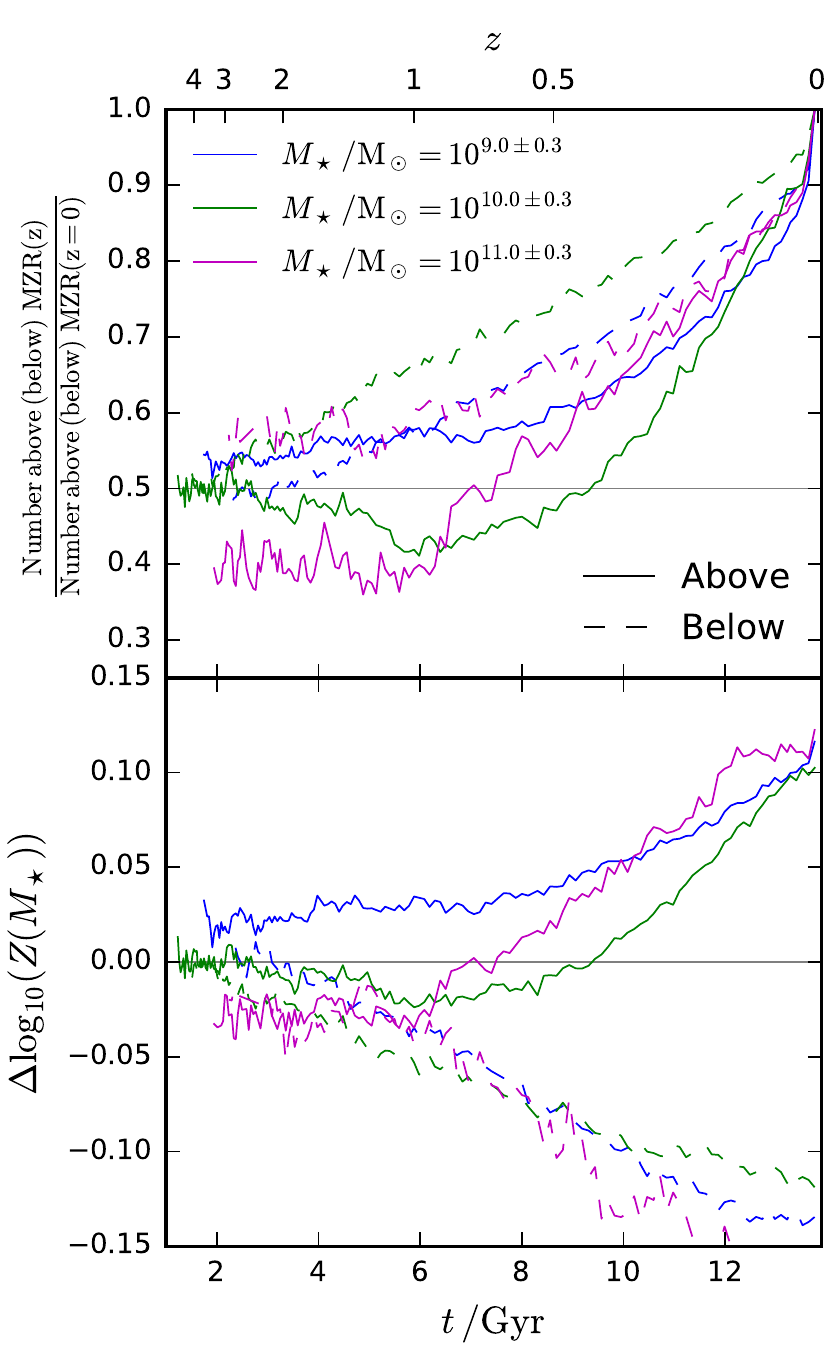}
\caption{\textit{Top:} The fraction of galaxies above (solid line style) or below (dashed line style) the MZR as a function of cosmic time, tracking the main progenitors of the 
sample of galaxies that are above or below the MZR at $z=0$ ($t = 13.8\, \mathrm{Gyr}$). The grey line at $y=0.5$ indicates what the fraction would be if the distribution were random within the samples at each redshift.
\textit{Bottom:} The median residual metallicity of galaxies in a sample above (solid line style) or below (dashed line style) the MZR at $z=0$ as a function of cosmic times. 
The grey line at $y=0$ indicates a random median residual at each redshift.
In both panels the stellar mass bins (blue: $M_\star/\mathrm{M}_\odot = 10^{9.0 \pm 0.3}$, 
green: $M_\star/\mathrm{M}_\odot = 10^{10.0 \pm 0.3}$, 
magenta: $M_\star/\mathrm{M}_\odot = 10^{11.0 \pm 0.3}$) refer to masses at $z=0$.
The sample above the MZR has been above the MZR for $\gtrsim 6 \, \mathrm{Gyr}$. The sample below the MZR has been below the MZR for $\approx 12 \, \mathrm{Gyr}$. 
}
\label{fig_rem_sc}
\end{figure}

We next consider the timescales over which galaxies remain above/below the MZR, independently of the other variables considered.
The top panel of Fig.~\ref{fig_rem_sc} shows the 'memory' of the scatter in the MZR by tracking a sample of galaxies that lie above (or below) the MZR at $z=0$ to higher redshifts. 
At each cosmic time the fraction of galaxies above (or below) the MZR at that time is determined, until a fraction smaller than 0.2 of the original galaxies remain. 
The bottom panel of Fig.~\ref{fig_rem_sc} shows instead the median residual metallicities of these galaxies at the different redshifts.

If the metallicity in the ISM of galaxies had no 'memory', a fraction of 0.5 (grey line, top panel) and a 
median residual of 0 (grey line, bottom panel) would be expected, as the distribution would be random at $z>0$. 
Fig.~\ref{fig_rem_sc} shows clearly that the metallicities of the galaxies do have a memory, as the scenario mentioned above does not hold. 

Galaxies above the MZR have on average been above the MZR for $\gtrsim 6 \, \mathrm{Gyr}$, if their current mass is $M_\star/\mathrm{M_\odot} \gtrsim 10^{10}$. 
Galaxies with present-day mass $M_\star/\mathrm{M}_\odot = 10^{9.0\pm0.3}$ have been above the MZR for on average $\approx12\, \mathrm{Gyr}$, i.e., their entire lifetime. 
Galaxies below the MZR have also been below the MZR for on average $\approx12\, \mathrm{Gyr}$. This idea is consistent with our hypothesis 
that processes like star formation, inflows and outflows determine the metallicity, since these also have trends over longer time scales 
(see figures \ref{fig_ind} and \ref{fig_tint}). Furthermore, the long timescales involved suggest that the metallicity is driven by the age 
of the halo formation itself. When comparing to the work by \cite{Matthee_2019} on the memory of the SFR - stellar mass sequence in \eagle, we see 
similar timescales ($\approx 10\, \mathrm{Gyr}$) appear, which they showed are connected to halo formation. Our results are also broadly
speaking consistent with the analysis of the relevant timescales for MZR and star formation rate evolution in the Illustris-TNG simulations, as
presented in  \cite{Torrey18}, who find that both fluctuate over long timescales.

The asymmetry between above and below the MZR may be due to the fact that galaxies below the MZR have a higher SFR, resulting in a more rapid decrease in stellar mass as we travel back in time, creating a difference between the two samples. Why this difference would result in a longer timescale is not immediately obvious however.

Together, Fig.~\ref{fig_tint} and Fig.~\ref{fig_rem_sc} show that the correlations between residual metallicity and residual specific inflow, 
outflow and star formation rate are consistent with being driven by fluctuations over longer timescales, coinciding with the typical 
timescales that galaxies have been above the MZR. An extension of our analysis would be to attempt to decompose the
fluctuation spectrum explicitly between different timescales, either using Fourier analysis \cite[e.g.][]{2019caplar,Enci20,Iyer20} or principal component
analysis \cite[][]{Matthee_2019}. We leave this as a possible avenue for future work.

\section{Summary}
\label{Summary}

We have presented our findings concerning the physical origin of the scatter in the Mass-Metallicity Relation (MZR) of central galaxies in the \eagle simulation and the roles of the gas fraction, specific inflow, outflow and star formation rates. By studying the residual correlations between ISM metallicity and these variables, comparing them, correcting them for one another and studying the timescales involved, it has become clear that although all of these variables are coupled, they also have an independent impact on the scatter in the MZR.
The gas fraction, which is most likely an aggregate of the other variables, is the variable correlating best with the scatter in the MZR and the inflow rate seems to play a larger role than the outflow or star formation rates.
This section summarises the main results.

\begin{enumerate}
\item{There are anti-correlations between the residuals of the MZR, and the residuals of the relations between stellar mass and, respectively, the gas fraction, specific inflow, outflow and star formation rate, at $M_\star/\mathrm{M}_\odot \lesssim 10^{10}$ (see Fig.~\ref{fig_mb}).
These residual correlations are comparable in shape for $M_\star/\mathrm{M}_\odot \lesssim 10^{10}$. For $M_\star/\mathrm{M_\odot} \gtrsim 10^{11}$ the shapes deviate compared to lower mass bins: the correlations between residual metallicity and residual sSFR and gas fraction change in sign, showing a positive correlation;
the correlations between residual metallicity and residual specific inflow and outflow rate also deviate in shape, showing a positive correlation at negative x-axis residuals, i.e. for relatively low flow rates, and an anti-correlation at positive x-axis residuals, i.e. for relatively high flow rates.
}
\item{The residual specific black hole mass (i.e. $\log_{10}(M_{\mathrm{BH}}(M_\star)/M_\star)$) anti-correlates with residual metallicity at $M_\star/\mathrm{M}_\odot \gtrsim 10^{11}$ and for lower stellar masses also for relatively high black hole masses (see Fig.~\ref{fig_mbh}). This suggests that at high black hole masses, AGN feedback strongly influences the gas metallicity of galaxies.
}
\item{The relations between the residuals of the MZR and the residual gas fraction, residual specific inflow, outflow and star formation rates flatten considerably when corrected for the dependence of metallicity on any of the other variables (see Fig.~\ref{fig_coup}). The relation between residual metallicity and residual specific star formation rate (sSFR) is completely determined by the gas fraction, and almost completely by specific inflow rate and specific outflow rate for non-extreme values ($-0.7 < \Delta \, \log_{10}(\mathrm{sSFR}) < 0.3$).
This suggests that all variables are coupled, but that they have some independent influence on the metallicity, of which the influence of the star formation rate is least strong. 
}
\item{When correcting the scatter in the MZR for the correlations between residual metallicity and residual specific inflow rate, residual specific outflow rate, residual specific star formation rate or the residual gas fraction, and measuring the remaining scatter, the scatter is reduced most (from $\approx0.13$ dex down to $\approx0.07$ dex) by the gas fraction, and similarly by the specific inflow rate, at $M_\star/\mathrm{M}_\odot <10^{10.2}$ (see Fig.~\ref{fig_expl_sc}). Accounting for the specific outflow or star formation rate reduces the scatter by $\approx0.025$ dex at $M_\star/\mathrm{M}_\odot <10^{9.3}$, compared to $\approx0.04$ dex at $M_\star/\mathrm{M}_\odot \approx 10^{9.6}$. At high stellar masses ($M_\star/\mathrm{M}_\odot >10^{10.2}$) correcting for the specific black hole mass reduces the scatter whereas the contribution of the other variables diminishes.
}
\item{When averaged over longer time intervals, the anti-correlations of the residual metallicity with the residual specific inflow and outflow rates steepen ($M_\star/\mathrm{M}_\odot = 10^{10.0\pm0.3}$, $z<0.1$), indicating that the fluctuations on longer timescales are responsible for the residual correlations (see Fig.~\ref{fig_tint}). The relation with the residual specific star formation rate ($M_\star/\mathrm{M}_\odot =10^{10.0\pm0.3}$, $z<0.1$) does not change when measured over longer time intervals, indicating that fluctuations on short timescales are not responsible for this residual correlation.
}
\item{When galaxies with $M_\star/\mathrm{M}_\odot \gtrsim 10^{10}$ above (below) the MZR at $z=0$ are tracked back in time, these have generally been above (below) the MZR for $\approx6$ or $\approx12\, \mathrm{Gyr}$, respectively (see Fig.~\ref{fig_rem_sc}). Lower-mass galaxies ($M_\star/\mathrm{M}_\odot = 10^{9.0\pm0.3}$) have generally been on one side of the MZR for $\approx12\, \mathrm{Gyr}$.
}
\end{enumerate}

Summarising, star formation, inflows and outflows of gas are coupled processes that all seem to have an independent impact on the metallicity of the ISM, 
of which the inflows have the most influence. 
Consistent with the analyis of the Illustris-TNG simulations presented in \cite{Torrey18}, our analysis of the \eagle simulations indicates that 
important timescales for these processes are of the order of a few $\mathrm{Gyr}$ or longer, 
which is reflected in how long galaxies stay above/below the MZR, on average.
There timescales are also consistent with the recent inference presented in \cite{Wang20}, based on the residuals measured in observational data.

Overall, we find that the gas fraction is still the variable that correlates best with the metallicity \cite[see also][]{Lagos16, De_Rossi_2017}. 
We expect that this is primarily because the gas fraction reflects the inflows, outflows and star formation over a range of timescales. 
This reinforces the general notion that the MZR and its scatter represents an important diagnostic for a range of processes that are critical for galaxy evolution,
but which are difficult to observe directly.

\section*{Acknowledgements}
This work is partly funded by Vici grants 639.043.409 from the Dutch Research Council (NWO). 

This work used the DiRAC@Durham facility managed by the Institute for Computational Cosmology on behalf of the STFC DiRAC HPC Facility (www.dirac.ac.uk). The equipment was funded by BEIS capital funding via STFC capital grants ST/K00042X/1, ST/P002293/1, ST/R002371/1 and ST/S002502/1, Durham University and STFC operations grant ST/R000832/1. DiRAC is part of the National e-Infrastructure.

\section*{Data availability}
The EAGLE simulation data are publicly available from 
http://icc.dur.ac.uk/Eagle/database.php, as described in \cite{2016McAlpine}. The processed data underlying this work will be shared on 
reasonable request to the corresponding author.

\bibliographystyle{mn2e}
\bibliography{Bibliography}
\appendix
\section{Convergence test}
\label{ap_recal}

We test the convergence of our results by comparing the Reference \eagle model at standard \eagle resolution against the re-calibrated Recal model at eight times higher mass
resolution \cite[][]{Schaye15}. To ensure a fair comparison in terms of sample sizes, we use $(25 \, \mathrm{Mpc})^3$ volume simulations for both models in this comparison, rather than the
$(100 \, \mathrm{Mpc})^3$ volume used in our main analysis using only the Reference model. The comparison is particularly salient in this study since the Recal simulation better reproduces the observationally inferred shape of the MZR at low galaxy stellar mass \cite[][]{Schaye15,De_Rossi_2017}.
Figure \ref{fig_mb_recal} shows the relations between the MZR residual and the residuals of the relations between stellar mass and, respectively,
specific inflow rate, specific outflow rate and specific star formation rate for different stellar mass bins 
($M_\star/\mathrm{M_\odot} = 10^{9.0 \pm 0.3}$ (blue), $M_\star/\mathrm{M_\odot} = 10^{10.0 \pm 0.3}$ (green), $M_\star/\mathrm{M_\odot} = 10^{11.0 \pm 0.3}$ (magenta)), at $z<0.1$.
Note that the number of galaxies in the data points for the $10^{11.0 \pm 0.3}$ stellar mass bin is very close to the cut-off of 10 galaxies, making the results for this stellar mass bin less robust.
The solid lines represent the lower resolution Reference model simulation, whereas the dashed lines represent the higher resolution Recal model simulation. Note that the median inflow rate in REF-L025N752 is slightly lower than in REF-L100N1504 for the $10^{9.0\pm 0.3}$ stellar mass bin, such that for REF-N025L752 only relatively high outflow/inflow/star formation rate residuals pass our 10-particle cut.
\\

The two simulations show the same qualitative trends but differ quantitatively in some respects. Given the smaller volume used for the comparison,
the Recal simulation shows a larger dynamic range towards lower residuals in specific star formation rate, inflow rate, and outflow rate, which
can be explained by the eight times smaller particle mass employed. The slopes of the correlations between residuals differ somewhat
in some cases, but the limited sample sizes available make it difficult to reach any firm conclusion.

\begin{figure*}
  \begin{center}
\includegraphics[width=40pc]{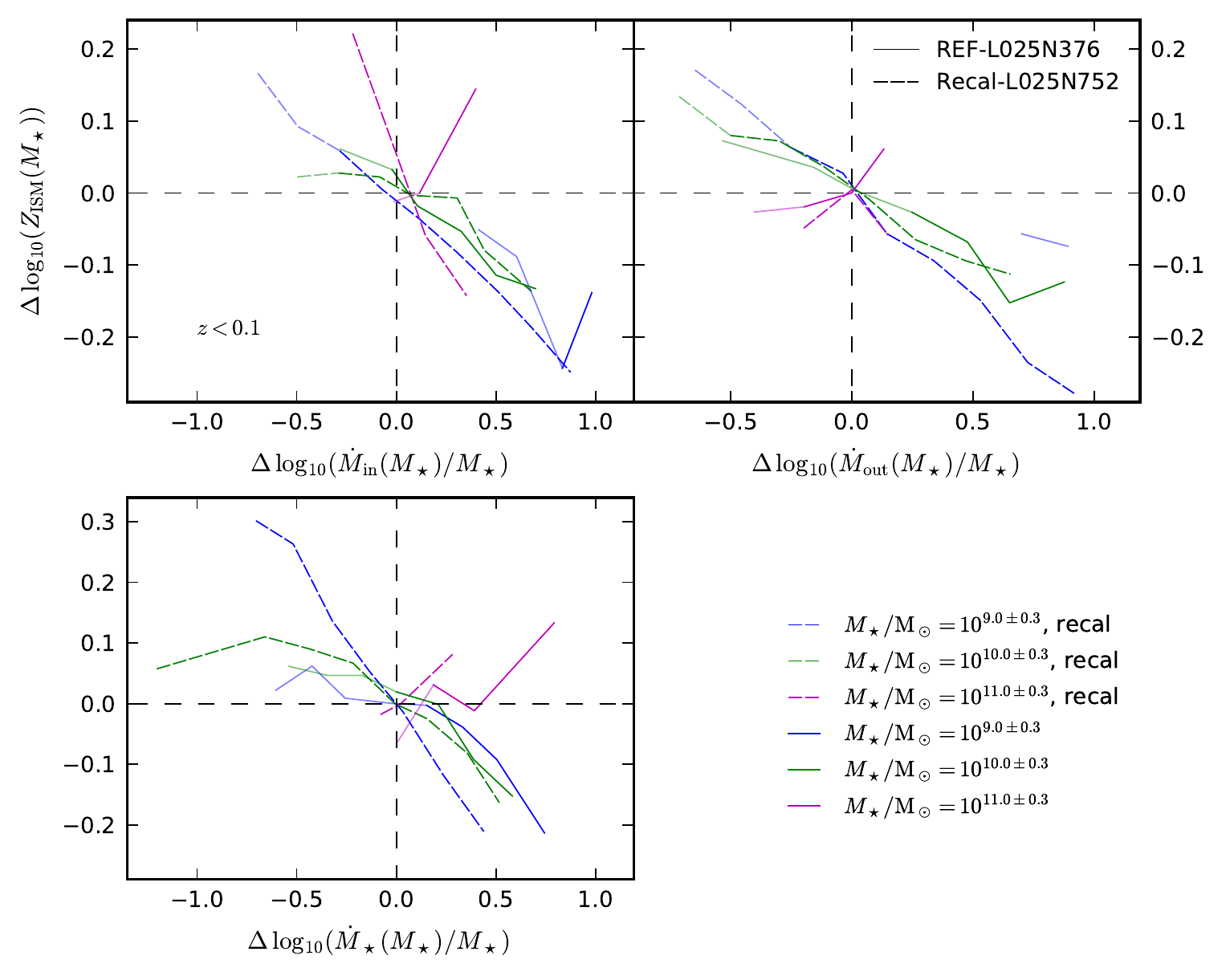}
\caption{The relations between the median residual metallicity of the interstellar medium and  residual specific inflow rate (top left), residual specific outflow rate (top right) and residual sSFR (bottom left) for central galaxies with $M_\star/\mathrm{M_\odot} = 10^{9.0 \pm 0.3}$ (blue lines), $M_\star/\mathrm{M_\odot} = 10^{10.0 \pm 0.3}$ (green lines), $M_\star/\mathrm{M_\odot} = 10^{11.0 \pm 0.3}$ (magenta lines), at $z<0.1$.
The residuals are taken with respect to the medians of each quantity as a function of stellar mass.
The dashed lines correspond to the higher resolution Recal simulation whereas the solid lines correspond to the standard resolution Reference model simulation.
Both simulations use a volume of $(25 \, \mathrm{Mpc})^3$, which is $64$ times smaller than the volume of the Reference model simulation used for our main analysis.
A transparent line style indicates regions where more than 50\% of galaxies in the bin contain $<$ 20 inflowing (top left panel), outflowing (top right panel) or newly formed star particles (bottom left panel).
The lines are cut off when more than 50\% of galaxies in the bin contain $<$ 10 inflowing (top left panel), outflowing (top right panel), newly formed star particles (bottom left panel) or when there are $<$ 10 galaxies in the bin.
Both simulations produce the same main qualitative trends, agreeing fairly well.}
\label{fig_mb_recal}
  \end{center}
\end{figure*}

\section{Redshift Evolution}
\label{Redshift_evolution}



The redshift evolution varies between the specific inflow rate, specific outflow rate, sSFR and gas fraction residual correlations (figure \ref{fig_redev}). For the calculations of the specific inflow, outflow and star formation rates different time intervals  were used for different redshifts, in approximation of 0.2 times the halo dynamical time, to account for the expansion of the universe.
Both the specific inflow rate and gas fraction residual correlations show an increase in steepness of the correlation at higher redshifts for galaxies with $M_\star/\mathrm{M_\odot} = 10^{10.0 \pm 0.3}$. Contrarily, the specific outflow residual correlation shows no clear redshift evolution trend and the sSFR residual correlation seems to flatten at higher redshift although there is no smooth progression. However, overal the redshift evolution is mild.



\begin{figure*}
  \begin{center}
\includegraphics[width=40pc]{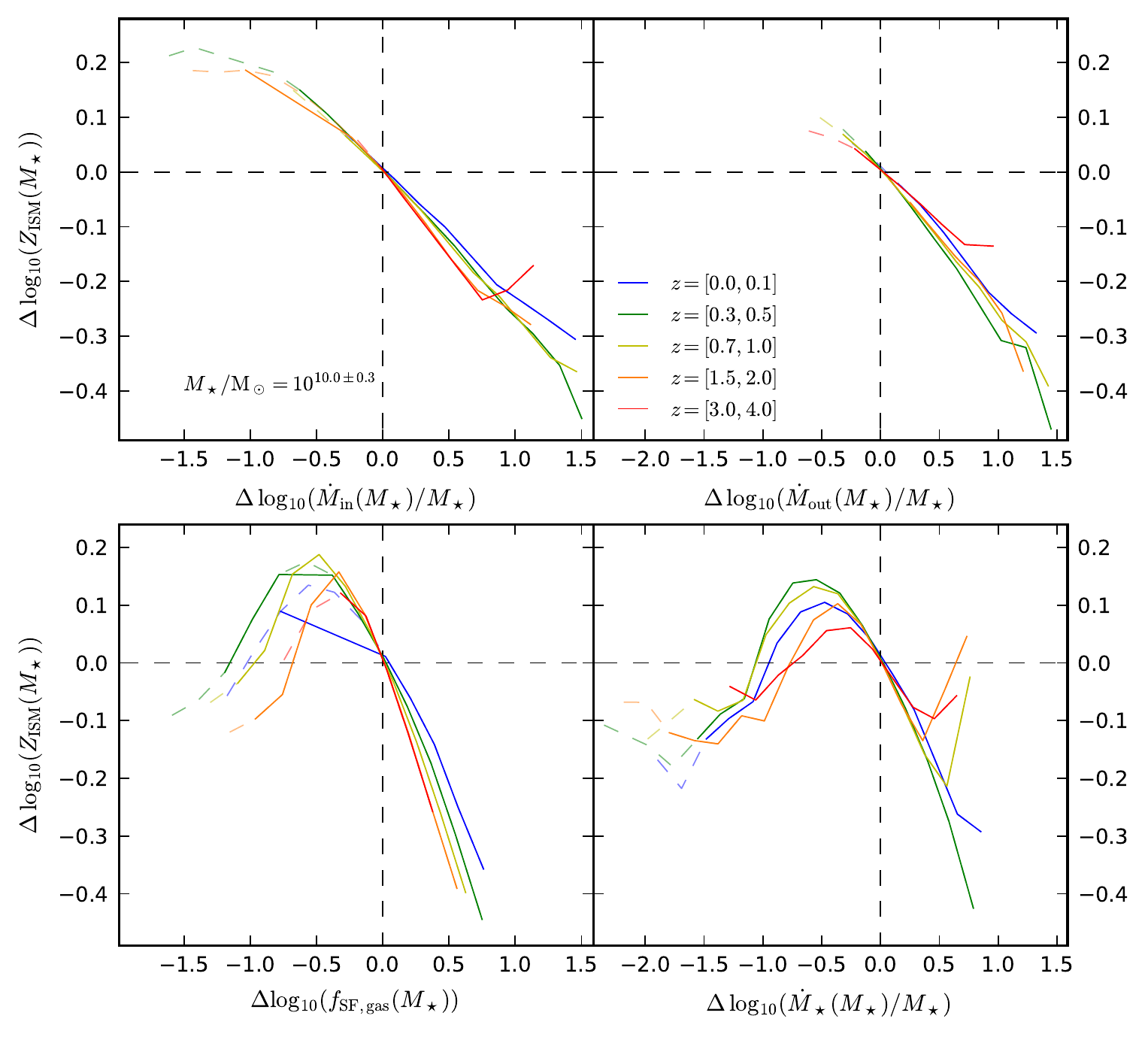}
\caption{The correlations between residual metallicity and residual specific inflow rate (top left), residual specific outflow rate (top right) residual gas fraction (bottom left) and residual sSFR (bottom right) for galaxies with $M_\star/\mathrm{M}_\odot = 10^{10.0 \pm 0.3}$, at different redshifts.
A time interval of 0.24 Gyr for $z<0.1$; 0.24 Gyr for $z=[0.3,0.5]$;  0.12 Gyr for $z=[0.7,1.0]$, 0.085 Gyr for $z=[1.5,2.0]$, 0.043 Gyr for $z=[3.0,4.0]$ between measurements is used, in approximation of 0.2 times the halo dynamical time at these redshifts.
Line styles are as in Fig.2.
The gas fraction and specific inflow rate residual correlations steepen with redshift while the sSFR residual correlation flattens with redshift. At $z>0.3$ the specific outflow rate residual correlation shows no clear redshift evolution. Overall, the evolution is weak.
}
\label{fig_redev}
  \end{center}
\end{figure*}


\label{lastpage}
\end{document}